# Mechanical and Energy-absorption Properties of Schwarzites

Levi C. Felix[1,2], Cristiano F. Woellner[3] and Douglas S. Galvao[1,2]*

[1]Applied Physics Department, State University of Campinas, Campinas/SP, 13083-970, Brazil

[2]Center for Computational Engineering & Sciences - CCES, State University of Campinas, Campinas/SP, 13083-970, Brazil.

[3]Physics Department, Federal University of Paraná - UFPR, Curitiba/PR, 81531-980, Brazil.

**Abstract**

We investigated through fully atomistic molecular dynamics simulations, the mechanical behavior (compressive and tensile) and energy absorption properties of two families (primitive (P688 and P8bal) and gyroid (G688 and G8bal)) of carbon-based schwarzites. Our results show that all schwarzites can be compressed (with almost total elastic recovery) without fracture to more than 50%, one of them can be even remarkably compressed up to 80%. One of the structures (G8bal) presents negative Poisson's ratio value (auxetic behavior). The crush force efficiency, the stroke efficiency and the specific energy absorption (SEA) values show that schwarzites can be effective energy absorber materials. Although the same level of deformation without fracture observed in the compressive case is not observed for the tensile case, it is still very high (30-40%). The fracture dynamics show extensive structural reconstructions with the formation of linear atomic chains (LACs).

*corresponding author: galvao@ifi.unicamp.br, Phone: +55-19-35215373

# 1. Introduction

Schwarzites are crystalline carbon allotrope structures with negative gaussian curvatures. These structures were proposed in 1991 by Mackay and Terrones [1]. They used the concept of negative curvature in the context of periodic graphitic structures, with the same shape as triply periodic minimal surfaces (TPMS). In other words, schwarzites consist of TPMS decorated with carbon atoms along the surface. The space group of Mackay and Terrones' schwarzites is $Im\bar{3}m$ (group number 229). A year later, Lenorsky *et al.* [2] demonstrated the stability of other schwarzite structures ($Pm\bar{3}m$, group number 221) which, in addition to hexagons and octagons also contain pentagons and heptagons. Further works demonstrated that larger stable Schwarzite structures can be build increasing the ratio of hexagons/octagons (heptagons), the so-called giant schwarzites [3,4].

In this work, we want to investigate how the ratio of hexagons/octagons affects the Schwartize mechanical response regarding compressive and tensile strains and how this determines the mechanical failure (fracture) patterns. Based on this, we selected four representative structures belonging to the gyroid and primitive schwarzite families (see Figure 1) [1,3–5]. It is important to stress that in carbon schwarzites planar (graphenic) and non-planar regions are both present and this affects the mechanical properties. As discussed by Miller *et al. [3]* the 'flat' regions are consequences of energy balance, since $sp^2$-hybridization favors planar structures like graphene. In this same work, they reported the appearance of ripples as the Schwarzite 'flatness' increases (increase of hexagons). This effect appears in order to thermodynamically stabilize graphene-like surfaces [6] in large flat (giant) Schwarzites.

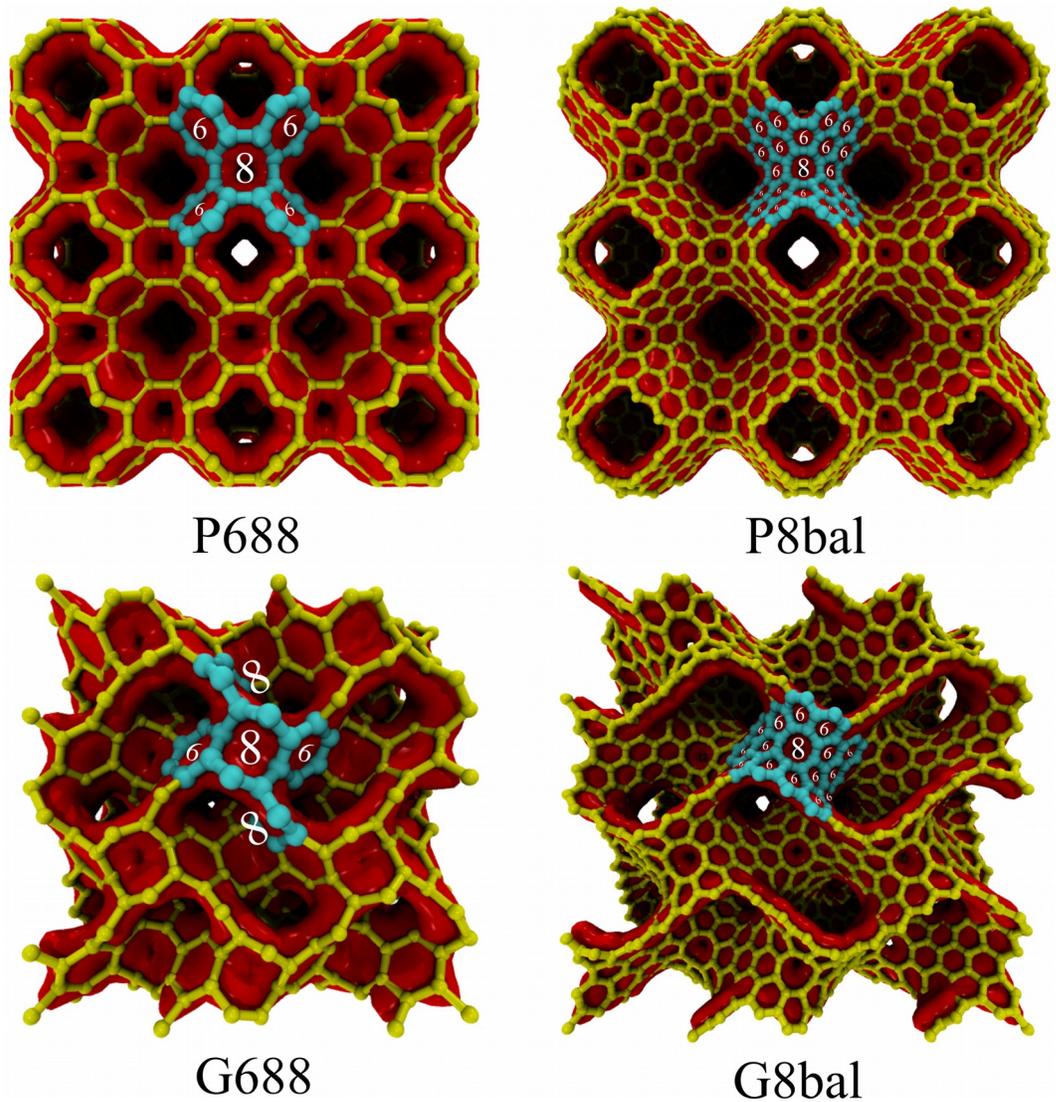

**Figure 1**: Supercells of four schwarzites from two distinct families discussed in this work. P688 and P8bal belong to the Primitive (P) family and G688 and G8bal belong to the Gyroid (G) family. The main difference between P8bal and G8bal with respect to their counterparts P688 and G688 is the higher ratio of hexagons to rings with more than six atoms, which makes them locally 'flatter'.

Other works investigated the stability of some schwarzite structures [7–10] and possible synthesis routes [11–14]. Although the complex geometry of schwarzites poses significant challenges for their synthesis, they have been used to model realistic synthesized carbon nanofoams [15–19] and to clarify the role of negative Gaussian curvature in chemical structures [20] and biophysical systems [21,22]. With the recent synthesis of new porous carbon networks [14,23] there is a renewed interest in schwarzite-like structures due to their unique topological properties and many potential applications, such as catalysis, molecular sieving [24,25], gas storage [26], alkali ion batteries [27], as anode for lithium-ion batteries [28] and energy-absorbing materials [15–19]. A recent work proposed a synthesis route to schwarzites through Zeolite templating [13], but up to now their synthesis remains elusive.

Regarding the electronic structure, the schwarzite band gap values vary from metallic to semiconductor behavior depending on the 'flatness' [5,29–32]. Interestingly, first-principles calculations [33] predict that structures can even exhibit semimetal Dirac-like points, like the ones present in graphene. Wide bandgap semiconductor and insulator behavior were recently predicted to boron nitride-based Schwarzites [34]. Schwarzites have also interesting magnetic properties, as some structures were predicted to present a net magnetic moment in their electronic ground state [35]. It has also been predicted that the presence of negative Gaussian curvatures can induce suppression in the lattice thermal conductivity [36,37], which can be further tuned by the introduction of guest atoms in their pores making them good candidates for thermoelectric applications [38].

We can expect that the unique schwarzite topology can be translated into interesting mechanical properties, in particular, the fracture patterns under compressive and tensile strains, have not yet been fully investigated in the literature. Molecular dynamics (MD) simulations have shown that they can stand high compressive loads without any fracture [3,39–41] and that the topology plays an important role in this process. This topology-related mechanical behavior has been exploited in 3D printed schwarzites at macroscale [42], where many features found at the atomic scale are also present in the macroscale. Even experimentally synthesized 3D graphene foams have shown to possess high compressive strength [17,43,44], making such feature inherent to many graphene-like porous structures.

In this work, we investigated the mechanical behavior under compressive and tensile strain of four different Schwarzites through fully atomistic reactive molecular dynamics (MD) simulations. We selected 4 representative structures of two schwarzite families (Gyroid and Primitive) with two structures into each family (see Figure 1). These chosen structures differ mainly through their local 'flatness' levels and they are representative of the large Schwarzite families. We also investigated their energy absorption behavior and contrasted them with other reference materials.

## 2. Materials and Methods

We carried out fully atomistic classical molecular dynamics (MD) simulations to investigate the mechanical properties and energy absorption behavior of selected schwarzite structures. The investigated structures belong to two schwarzite distinct families, named Gyroid and Primitive [3,4]. For the Gyroid family, we considered two different structures, the G688 and G8bal and for the Primitive family, the P688 and P8bal ones (Figure 1). The structures that belong to the same family differ mainly through their local 'flatness' (ratio of the number of hexagons to octagons). In the primitive family, for example, the smallest number of hexagons separating two octagons is one for the P8bal and zero for P688.

All MD calculations were carried out using the reactive force field ReaxFF [45], as implemented in the open-source code Large-scale Atomic/Molecular Massively Parallel Simulator (LAMMPS) [46]. ReaxFF is a widely used potential in molecular dynamics simulations, it allows the investigation of chemical processes, such as formation and breaking of chemical bonds. It has been largely used in different studies of carbon-based structures, such as high-velocity impact structures on graphene [47,48], graphene healing mechanism [49,50], curved carbon nanostructures [51], among others. ReaxFF parameterization is obtained from accurate calculations using Density Functional Theory (DFT) and/or when available, from experimental data [50]. Due to its reduced computational cost, ReaxFF allows carrying out long-time simulations and to handle large systems that are prohibitive or even impossible to treat using *ab initio* methods.

Snapshots of structures and movies were generated using the Open Visualization Tool (OVITO) [52].

Cyclic boundary conditions along all directions were used in all MD simulations. For the compressive/tensile strain simulations, a thermalization process is performed before deforming the structures. This thermalization process is important to eliminate any residual stress accumulation in the structures due to thermal effects. Thus, all structures are first thermalized under NPT ensemble controlled by a Nosé/Hoover chain of thermostats and a barostat. The temperature was fixed at 10 K and the fixed applied external pressure set to zero along all periodic directions. The thermalization process was carried out during 100 ps, which is enough for all structures to reach a relaxed state, *i.e.*, no more time large oscillations are observed in pressure and temperature values. This process is then followed by the structural deformations (compression/stretching, through a gradual decrease/increase of the crystal dimensions). All simulations were performed using a time step of 0.1 fs.

In order to allow a more direct comparison for all structures, we carry the MD simulations using simulation cells with cubic symmetry, which provides a simpler analysis of the elasticity tensor, as it reduces the number of independent matrix elements to three. Also, in order to avoid spurious effects due to size effects, in our simulations we used 3x3x3 supercells for all structures, except for the P688 structure where we used a 4x4x4 supercell because it is the smallest structure (it has only 48 atoms in the cubic unit cell). In Table 1 we present structural information, alongside with the number of atoms for each structure.

| Structure | n | N | a (Å) | ρ (g/cm³) |
|---|---|---|---|---|
| P688 (P8-0) | 48 | 3072 | 7.80 | 2.02 |
| P8bal (P8-1) | 192 | 5184 | 14.90 | 1.15 |
| G688 (G8-0) | 96 | 2592 | 9.60 | 2.16 |
| G8bal (G8-1) | 384 | 3072 | 18.41 | 1.23 |

**Table 1**: Structural information for all four schwarzites considered in this work. The columns represent, respectively: the number of atoms per cubic unit cell (n) and the total number in the supercell (N), lattice parameter (a) and mass density (ρ). In the first column, the names into parentheses refer to alternative nomenclature used in other works [3]. The first letter indicates the family of TPMS, which is followed by a number 8 indicating the introduction of octagons to produce the corresponding negative Gaussian curvature. Then, the number of minimum hexagons separating two octagons is given at the end.

The compression/stretching process employed in this work consists of a uniaxial load along a high symmetry direction, which was simulated through a gradual decrease/increase of the lattice parameter along the respective direction. We set a constant strain rate of $10^{-5}$ fs$^{-1}$ along the Z ([001]) direction for all simulations. This procedure was repeated for another 100 ps, which is enough to observe fracture for all structures.

In this way, the strain was defined as

$$\varepsilon_L = \frac{|L-L_0|}{L_0} \qquad (1)$$

where $L$ and $L_0$ are the deformed and the initial length of the structures along the compression/stretching direction, respectively. From molecular dynamics, the mechanical response can be obtained by calculating the virial stress tensor, defined as [53]

$$\sigma_{ij} = \frac{\sum_k^N m_k v_{ki} v_{kj}}{V} + \frac{\sum_k^N r_{ki} f_{kj}}{V} \qquad (2)$$

where the first and the second terms are respectively the kinetic and pairwise energy (virial term) contributions of all $N$ atoms, and $V$ is the supercell volume.

Regarding the mechanical response of materials, we usually have an elastic regime and then a plastic regime up to failure. The elastic regime is normally identified as a linear portion of the compressive/tensile stress-strain curves for low values of strain. However, some previous works [54,55] on the tensile deformation of graphene have shown that its elastic response is nonlinear. The aforementioned works employed a quadratic dependence of the stress on strain, given by

$$\sigma_L = E\varepsilon_L \qquad (3)$$

where $\sigma_L$ is the virial stress component along the uniaxial deformation direction, $\varepsilon_L$ is the longitudinal strain and $E$ is the Young's modulus. We, then, obtain an estimation for the Young's modulus from molecular dynamics simulations by performing a linear fitting in the elastic regime, which is usually the low-strain region. Schwarzites can present nonlinear behavior in the low-strain regime also due to its foam-like structure [56].

Another important parameter in the elastic regime is the Poisson's ratio $\nu$, defined as the negative value of the ratio between the transverse deformation $\varepsilon_T$ and the longitudinal one $\varepsilon_L$ (strain direction),

$$\nu = -\frac{\varepsilon_T}{\varepsilon_L} \qquad (4)$$

Unlike the stress-strain behavior, strain-strain behavior presents a well defined linear regime where the Poisson's ratio values were estimated by a linear fitting (see Supplementary Information for details in the Figures S1-S4).

In order to have an estimation of the spatial stress distribution during the compression/stretching process, we have also calculated the scalar quantity called von Mises stress [57] per atom $k$, defined as:

$$\sigma^k = \sqrt{\frac{(\sigma_{XX}^k - \sigma_{YY}^k)^2 + (\sigma_{YY}^k - \sigma_{ZZ}^k)^2 + (\sigma_{XX}^k - \sigma_{ZZ}^k)^2 + 6((\sigma_{XY}^k)^2 + (\sigma_{YZ}^k)^2 + (\sigma_{ZX}^k)^2)}{2}} \qquad (5)$$

The components $\sigma_{XX}$, $\sigma_{YY}$, and $\sigma_{ZZ}$ are the normal stresses, while $\sigma_{XY}$, $\sigma_{ZX}$, and $\sigma_{YZ}$ are the shear stresses. From the von Mises stress values it is possible to investigate the stress distribution along with the structure and also the critical behavior near the structural failure (fracture).

The energy absorption during the compression process can be evaluated using the *crush force efficiency*, the *stroke efficiency* and specific *energy absorption*.

The crush force efficiency, η, is given by:

$$\eta(\varepsilon) = \frac{\langle\sigma(\varepsilon)\rangle}{\sigma_{max}(\varepsilon)} \qquad (6)$$

where <σ(ε)> is the average stress up to strain ε and σ_max(ε) is the maximum stress value up to strain ε. The stroke efficiency, or densification strain, is defined as the strain value that maximizes the *energy absorption efficiency*, η_t(ε), given by:

$$\eta_t(\varepsilon) = \frac{1}{\sigma(\varepsilon)} \int_0^\varepsilon \sigma(\varepsilon') d\varepsilon' \qquad (7)$$

where σ is the stress and ε is the strain.

The specific energy absorption is given by:

$$\frac{1}{\rho} \int_0^\varepsilon \sigma(\varepsilon') d\varepsilon' \qquad (8)$$

where ρ, ε, and σ are the structural density, strain and stress, respectively. This energy is usually calculated in the densification strain. As will be shown later, however, the densification strain in one of the structures exceeds the fracture strain of the others. Thus, for a better comparison, all the energy absorption parameters will be evaluated at 50% strain, which is a high enough deformation, but before the fracture for all four structures.

## 3. Results and Discussion

In Figure 2 we present the compressive stress-strain curves for all four structures under uniaxial compression along the [001] direction, up to the failure strain ($\varepsilon_F$). In Table 2 we present a summary of the mechanical properties for the compressive cases.

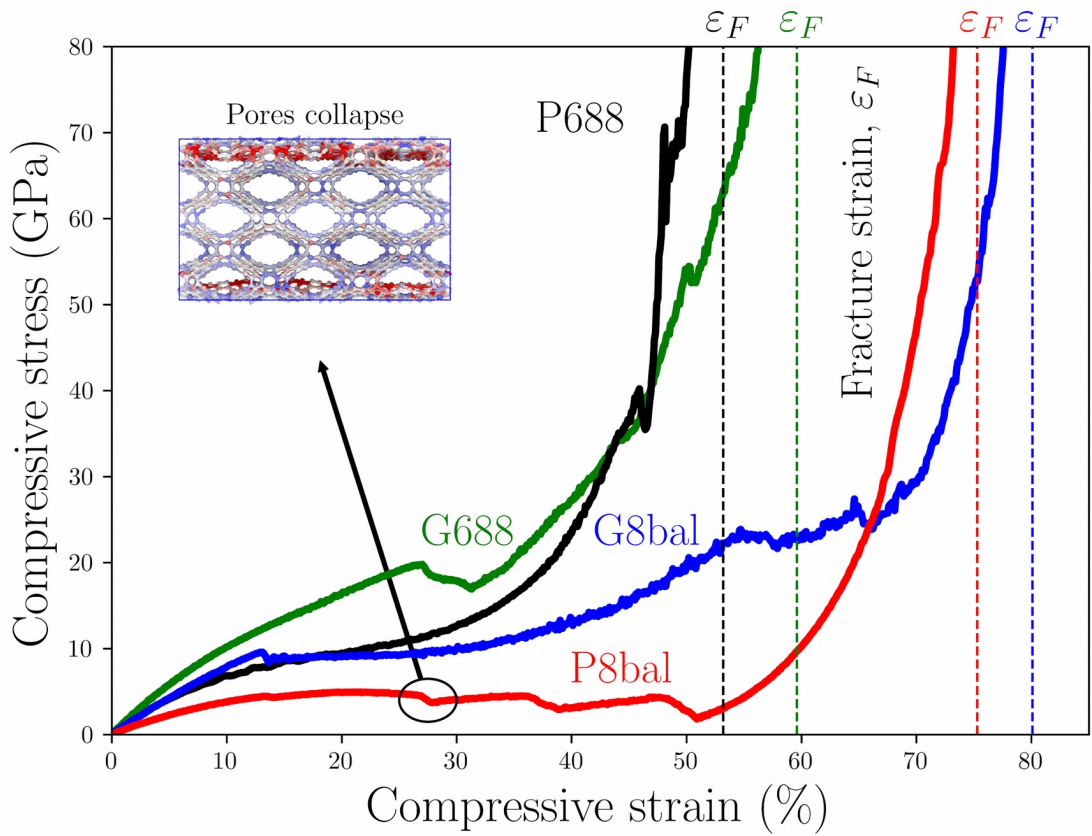

**Figure 2**: Compressive stress-strain curves for the four schwarzites structures under uniaxial compression along the [001] direction up to the fracture strain ($\varepsilon_F$). Notice that for each family (P and G) the higher the ratio of hexagonal rings to octagonal rings (P688 to P8bal and G688 to G8bal) the more resistant to failure the structure becomes, whereas the stiffness (Young's modulus) decreases. The inset shows the collapse of a set of pores in a P8bal structure, which is responsible for the plateaus observed in the red curve. The colors on the atoms represent the von Mises stress values at each atom from lower (blue) to higher (red) values.

| Structure | Young's modulus (E) [GPa] | Poisson's ratio (ν) | Fracture strain ($\varepsilon_F$) [%] | Compressive strength [GPa] |
|---|---|---|---|---|
| P688 (P8-0) | 86.98 | 0.45 | 53.18 | 162.69 |
| P8bal (P8-1) | 48.25 | 0.42 | 75.29 | 137.13 |
| G688 (G8-0) | 124.07 | 0.43 | 59.60 | 328.50 |
| G8bal (G8-1) | 85.23 | 0.35 | 80.09 | 194.74 |

**Table 2**: Calculated mechanical properties for the four schwarzites under compression. The columns represent, respectively: estimated Young's modulus value *E*, Poisson ratio ν, fracture strain, and compressive strength (at the fracture strain).

During these compressive loadings we observed three distinct regions in the stress-strain curves: (i) an elastic regime, where the stress increases (approximately) quadratically with the increase of compressive strain; (ii) a collapse plateau, corresponding to the pore buckling and/or collapsing and, finally, (iii) a densification regime, in which the stress rapidly increases up to the compressive strength. These behaviors are typical of foam-like structures [56].

From Figure 2, comparing the structures within the same family [(P688 and P8bal) and (G688 and G8bal)] we can recognize a common behavior: greater ratio values of hexagons to octagonal rings result in structures more resistant to fracture. It means that the structures can stand higher loads before collapsing. For both families, the increase in the fracture strain is over 20% (see details in Table 2). Unlike the fracture strain, the "flatter" a structure is within a given family (P or G) the lower its Young's modulus values, as can be seen in Table 2. This trend also holds for the collapse plateau as the structures buckle in lower stress for a decreasing local curvature ("flatter"). This feature is desirable for applications in energy-absorbing materials. It can also be observed some anomalies in the stress-strain curves in Figure 2, such as the three kinks for the P8bal structure. They correspond to the collapse of a set of pores (as shown in the inset of Figure 2). We emphasize that the collapse mechanism differs among different structure topologies [56]. As schwarzites are porous and very elastic structures, it is expected that the Young's modulus values would not be high. In fact, for the highest case (G688, 124.07 GPa) it is about 10 % of the graphene value [54].

Figure 3 presents the Schwarzites structures in the densification regime before the complete failure with lateral and top views for each structure. The color of the atoms and bonds represents the local stress (von Mises stress) and goes from blue (low stress) to red (high stress). We can notice that while for P8bal and G8bal, the stress is accumulated more or less evenly distributed, for P688 and G688 the stress is concentrated in specific regions. These features have implications in the structural elastic recovery discussed below. The lateral views show that all structures undergo densification through accordion-like mechanisms (the final stages resemble stackings of 2D graphene-like structures). But unlike typical van der Waals 2D structures (such as graphene and transition metal dichalcogenides) in this case, our '2D structures' remain covalently bonded. The whole process of the compression dynamics can be better understood through Videos 01 and 02 in the supplementary materials.

The results presented in Figures 2 and 3 and Table 2 show that schwarzites are remarkable structures from the point of view of standing very large deformations without structural fractures. All structures present fracture strain higher than 50% and in the case of G8bal exceptional 80.09%. Although large deformations without fracture are common among amorphous structures (such as foams), this is exceptionally rare for defectless crystalline structures, which make schwarzites unique structures.

Figure 4 shows the transverse (perpendicular direction) strain variation as a function of the applied compressive strain. The first derivative of these curves gives us the Poisson ratio ($\nu$). The values of the Poisson ratio (described in the previous section) in the linear regime are presented in Table 2 and are consistent with other reported works [3,39].

Remarkably, from Figure 4 we can see that one of the structures, the G8bal, presents negative Poisson's ratio, which means that the transverse direction shrinks as the load increases (see also Figure S5 in supplementary materials). Materials with this property are called Auxetic materials. Auxeticity for carbon-based materials have been predicted [58] and experimentally observed [59]. Interestingly, this behavior is only at certain strain range, but extensive one, from 13% up to 34%. This behavior can be attributed to its re-entrant topology under compression (see Videos 01 and 02) in supplementary materials), which is present in other auxetic materials [56,60].

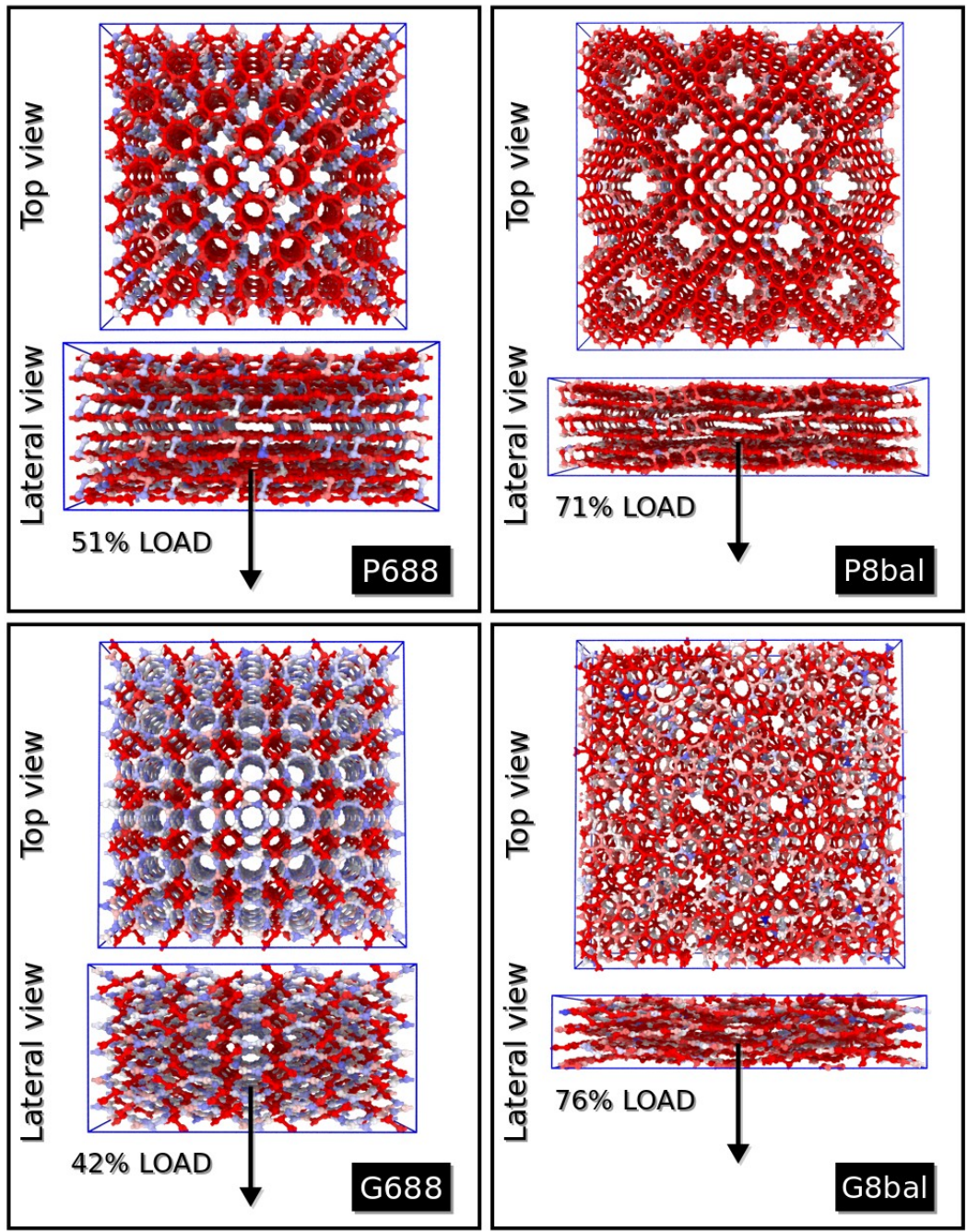

**Figure 3**: Representative MD snapshots for all schwarzites structures just before and after the failure strain ($\varepsilon_F$). The different color represents the local stress from low (blue) to high (red) stress value. Notice the lateral stacking-like densification present in all structures.

We also investigated the schwarzite elastic recovery under load/unload cycles. In Figure 5 we present the results for compressive strain up to 50% (see MD snapshots in Figure S6 of supplementary materials) and unloadings up to zero stress. We applied a uniaxial compression along the [001] direction up to 50% strain followed by unloading up to zero strain values.

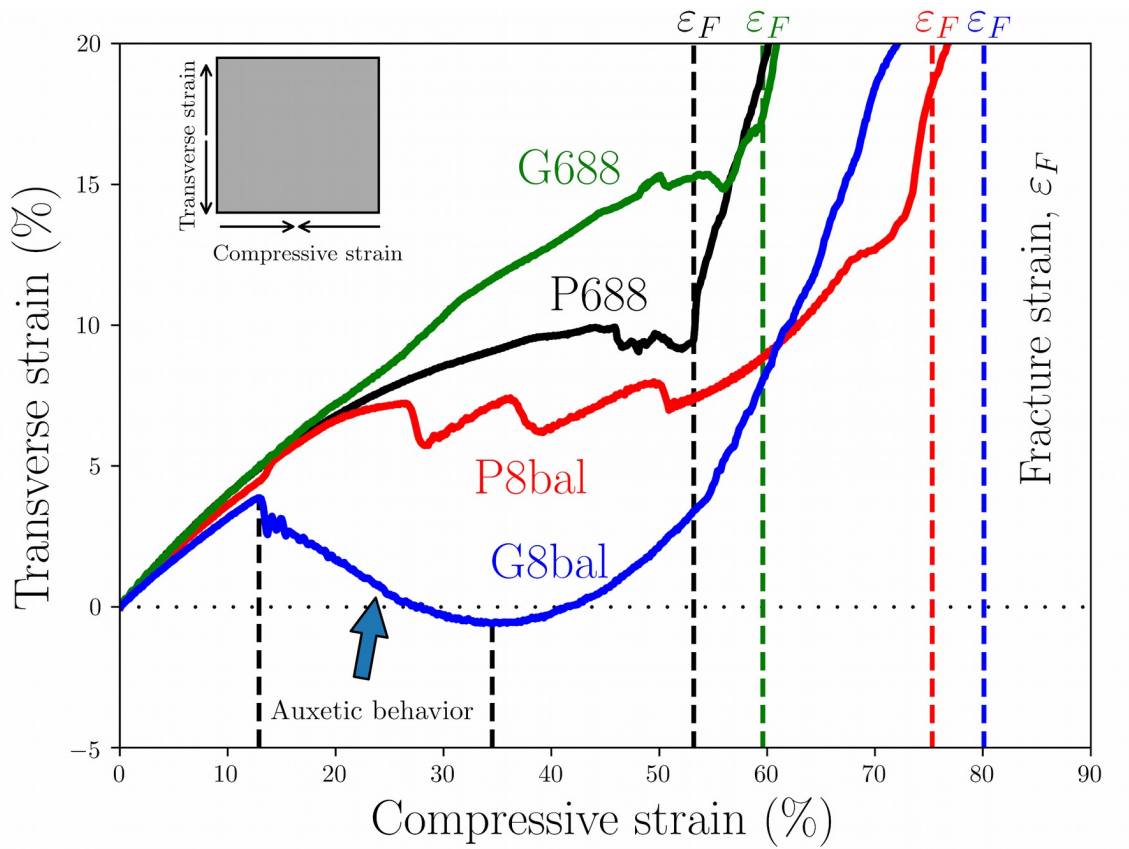

**Figure** 4: Transverse strain variation as a function of the applied compressive strain. The G8bal structure presents an auxetic behavior for strains between 13% and 34%, in which a structure expands/contracts in the direction perpendicular to the tensile/compressive stress.

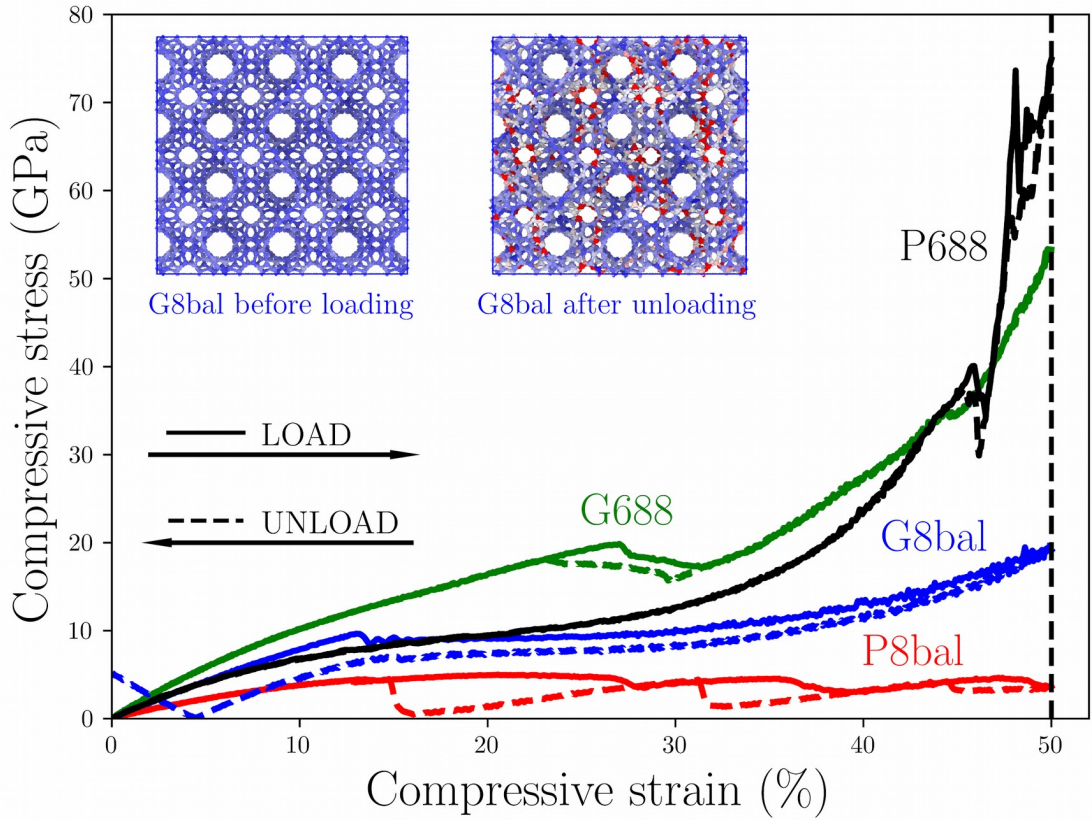

**Figure** 5: Compressive (load/unload) stress-strain curves for the four schwarzite structures under uniaxial compression along the [001] direction up to 50% and their unloading up to zero stress. The inset shows the von Mises stress distribution for G8bal structure before loading and after unloading (both at zero strain). The colors represent the von Mises stress regimes from low (blue) to high (red).

Unlike what is typically seen in foam-like structures, in spite of partial hysteresis, with the exception of G8bal all the other structures present an almost complete structural recovery, i.e. load (solid line) and unload (dashed line) curves are overlapped.

For the G8bal structure, the hysteresis comes from the pore collapse during loading, which results in irreversible structural changes that preclude the total elastic recovery. The final configuration (see insets of Figure 5, where can see changes in the pore shape) is a structure with residual and spatially localized stress (see Video 03). As expected, if this configuration is thermally annealed it goes back to its initial minimum energy configuration. For the other structures although these structural changes are also present the barriers for interconversion are low, which allows their almost total elastic recovery.

The patterns present in Figures 2 and 5 show similarities with materials that are good energy absorbers [19]. Based on that we calculated some magnitudes that characterize this feature. The results are presented in Table 3. We computed the values for the crush force efficiency, the stroke efficiency and the specific energy absorption (SEA).

SEA is one important parameter to characterize material performance. The SEA values we obtained are orders of magnitude higher than Kevlar and some steels and even higher than those recently reported for other carbon nanostructures [19] considered as very attractive candidates for energy absorption applications. The crush force and the stroke efficiency values are also consistent with a high-performance material. Considering the results presented in Table 3, schwarzites can be considered extremely attractive for any application in which lightweight yet highly energy absorbent materials are needed, such as body armor. There have been important synthesis advances for schwarzites inside zeolites [11,13,61], and large-size structures could be a reality in the coming years. Also, a recently 3D printed version of atomic schwarzite models was reported [42] and surprisingly some of their mechanical properties proved to be scale-independent. This suggests that some of the data and/or conclusions obtained here can be used to optimize or design better 3D printed structures.

| Structure | Crush force efficiency ($\eta$) | Stroke efficiency ($S_E$) [%] | Specific energy absorption at 50% for compression (MJ kg$^{-1}$) | Specific energy absorption at 50% for loading-unloading cycle (MJ kg$^{-1}$) |
|---|---|---|---|---|
| P688 (P8-0) | 0.22 | 0.33 | 4.03 | 0.11 |
| P8bal (P8-1) | 0.77 | 0.51 | 1.67 | 0.50 |
| G688 (G8-0) | 0.36 | 0.31 | 4.42 | 0.07 |
| G8bal (G8-1) | 0.51 | 0.66 | 4.03 | 0.69 |

**Table 3**: Calculated energy absorption properties for all schwarzites under compression. The columns represent, respectively: crush force efficiency, stroke efficiency, specific energy absorption calculated at 50% strain under uniaxial compression along the Z direction and specific energy absorption calculated at the 50% strain under loading-unloading cycle, also along the Z direction.

We have also considered tensile stress-strain analyses up to the limit of 100% strain, where all structures are already fractured. In Table 4 we summarize the results. In Figure 6 we present the tensile stress-strain curves for the four schwarzites under uniaxial tensile stress along the [001]. Similarly to the compressive case, we estimated Young's modulus values from the linear part of the curves displayed in Figure 6. Although they are of the same order, all values are higher than the corresponding ones in the compressive case. This indicates that it is easier to compress than to stretch the structures. The difference in Young's modulus values for compression/stretching is a common feature in foam-like materials [39,56].

| Structure | Young's modulus (E) [GPa] | Poisson's ratio (ν) | Fracture strain ($\varepsilon_F$) [%] | Tensile strength [GPa] |
|---|---|---|---|---|
| P688 (P8-0) | 118.64 | 0.48 | 41.70 | 136.47 |
| P8bal (P8-1) | 67.53 | 0.45 | 44.30 | 68.11 |
| G688 (G8-0) | 147.76 | 0.47 | 31.80 | 115.15 |
| G8bal (G8-1) | 127.79 | 0.44 | 32.30 | 49.64 |

Table 4: Calculated mechanical properties for all four schwarzites under stretching. The columns represent, respectively: Young's modulus E, Poisson's ratio ν, fracture strain $\varepsilon_F$ and Tensile strength (at the fracture strain).

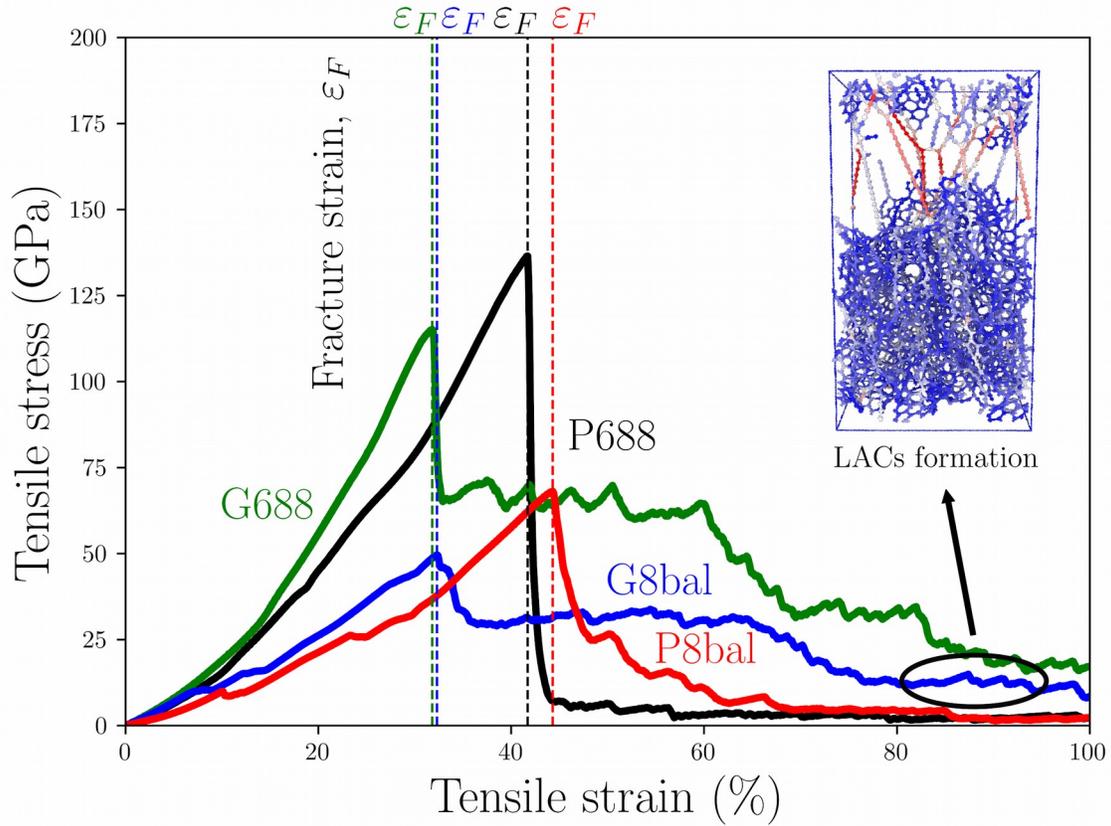

**Figure 6**: Tensile stress-strain curves of the four schwarzites under uniaxial stretch along the [001] direction up to 100% strain. Notice the presence of the formation of linear atomic chains (LACs) at the final stage (shown as inset for the G8bal structure). Colors in the inset represents the von Mises stress regime, from low (blue) to high (red).

In Figure 6, the abrupt drop in the tensile stress values is associated with structural failure (fracture), the so-called fracture strain ($\varepsilon_F$). They are indicated by dashed vertical lines in Figure 6. These values (see Table 4) are below 50% for all cases, in contrast to the compressive case, where it is possible to compress up to 80%. This again reflects the fact that is easier to compress than to stretch the structures. For the same reason, the tensile strength values are much lower than the corresponding compressive strength ones.

Before and after the fracture we can observe many changes in the curve slopes, which are indicative of structural changes. These changes are more pronounced after the fracture and are associated with the formation of linear atomic chains (LAC) that are responsible for the observed non-zero stress (in a saw-like shape). The inset shows one example of these reconstructions for the G8bal at 90% strain. LAC formation is a common feature in the simulations of fractured carbon-based materials, such as carbyne [62], novamene [63] and protomene [64], etc., and have been experimentally observed [65–68]. The structures of the P family presents a brittle behavior, which was also reported by Jung and Buehler [39] for similar structures. This brittle behavior occurs due to the sharp local stress distribution, as can be seen in Figure 7. The fracture dynamics are different for different families. The fractures occur mainly along orthogonal (P)/diagonal (G) planes to the stretching direction (see Video 04 and Video 05 in the supplementary materials).

In Figure 8 we present the strain variation in one of the perpendicular (transverse) directions, as a function of the applied tensile strain. Up to the fracture strain, both perpendicular directions have the same strain variation. The Poisson's ratio values are almost the same for all structures (see Table 4). However, in contrast to the compression case, auxetic behavior is no longer observed. This can be attributed that the stretching process precludes the re-entrant behavior, that is the origin of the auxeticity. Also, unlike the compression case (Figure 2), a higher number of hexagons in the structures (P8bal and G8bal) does not significantly increase the structural resistance to failure ($\varepsilon_F$). For instance, the structures G688 and G8bal (same family) have almost the same fracture strain ($\varepsilon_F$). In the elastic regime, a higher number of hexagons decreases the stiffness (Young's modulus value).

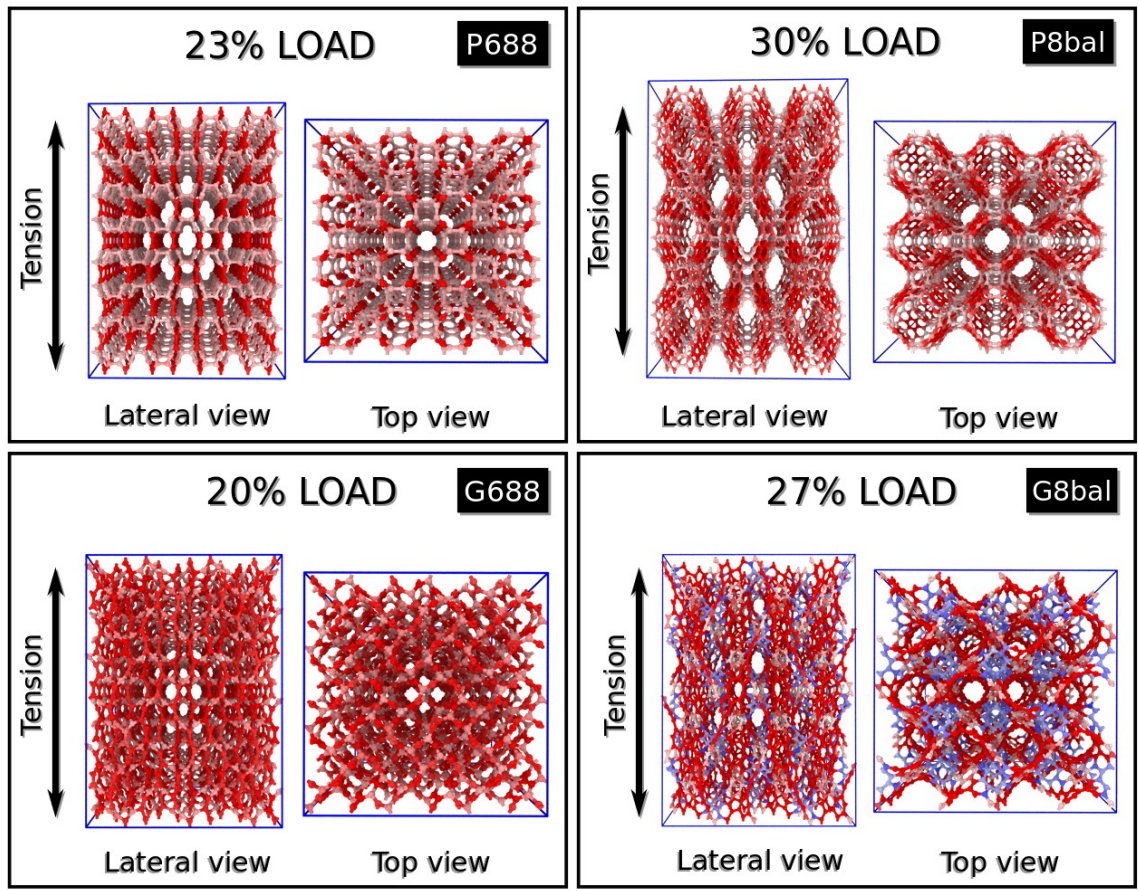

**Figure 7**: Representative snapshots from molecular dynamics simulations of all four schwarzite structures right before the fracture strain. The color represents the local stress and varies from low (blue) to high (red) stress.

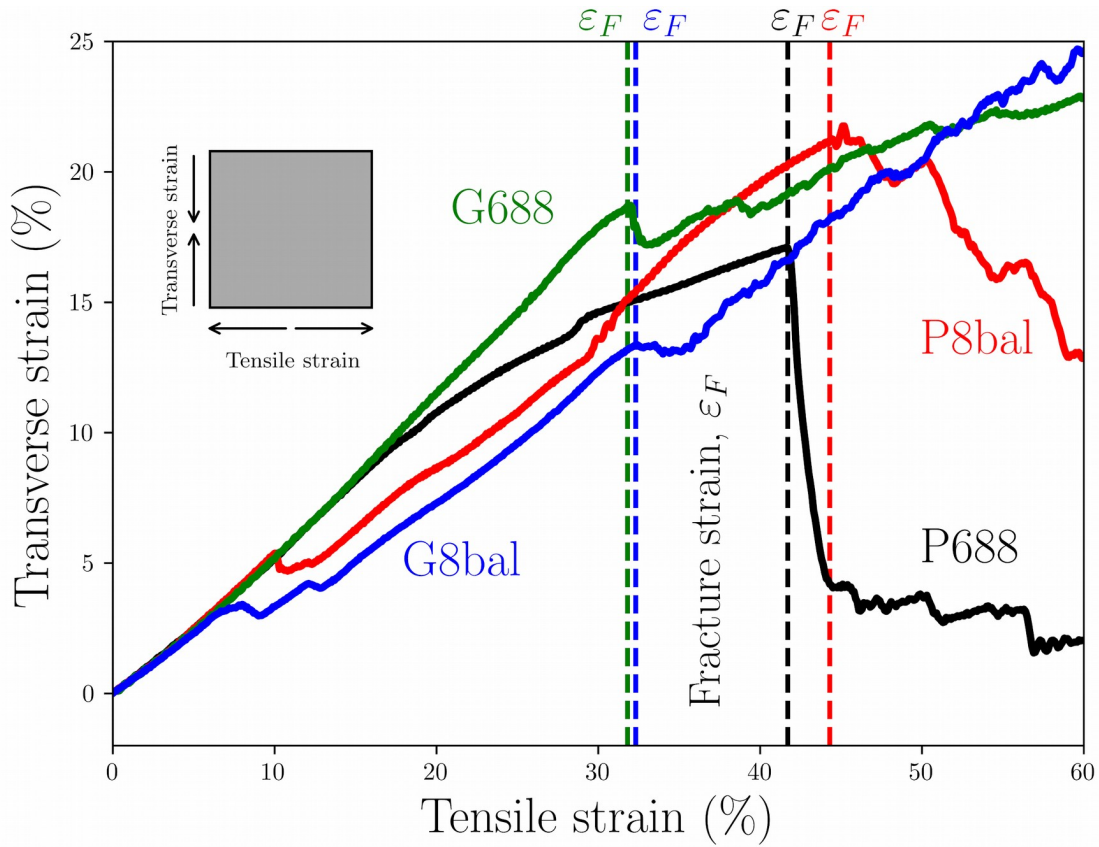

**Figure** 8: Transverse strain variation as a function of the applied tensile strain up to 60% strain. Both perpendicular directions have the same strain variation up to the fracture strain.

## 4. Summary and Conclusions

We investigated through fully atomistic molecular dynamics simulations, the mechanical behavior (compressive and tensile) and energy absorption properties of two families (primitive (P688 and P8bal) and gyroid (G688 and G8bal) of carbon-based schwarzites, which are crystalline porous structures with triply-periodic minimal surfaces (negative curvatures).

Our results show that all schwarzites can be compressed without fracture to more than 50%, one of them can be even remarkably compressed up to 80%, with typical Young's modulus values ~ 10% of graphene. This behavior can be explained by the presence of large pores and a re-entrant mechanism. One of the structures (G8bal) presents negative Poisson's ratio value (auxetic behavior) in the strain range from 13 up to 34%. Also, except for G8bal, although there are intermediate hysteresis, all structures present total elastic recovery from 50% load/unload cycles. Although large deformations without fracture are common among amorphous structures (such as foams), this is exceptionally rare for defectless crystalline structures, which make schwarzites unique structures.

The crush force efficiency, the stroke efficiency, and the specific energy absorption (SEA) values show that schwarzites can be effective energy absorber materials. The SEA values are orders of magnitude higher than Kevlar and some steels and even higher than those recently reported for other carbon nanostructures [19].

For the tensile stress, although Young's modulus values are of the same order, all values are higher than the corresponding ones in the compressive case. This indicates that it is easier to compress than to stretch the structures. The auxetic behavior is no longer observed. Although the same level of deformation without fracture observed in the compressive case is not observed, it is still very high (30-40%). The fracture dynamics show extensive structural reconstructions with the formation of linear atomic chains (LACs).

There have been important synthesis advances for schwarzites inside zeolites

[11,13,61], and large-size structures could be a reality in the coming years. Also, a recently 3D printed version of atomic schwarzite models was reported [42] and surprisingly some of their mechanical properties proved to be scale-independent. This suggests that some of the data and/or conclusions obtained here can be used to optimize or design better 3D printed structures. We hope the present work could stimulate further studies on these remarkable carbon allotrope structures.

**Acknowledgements**

This work was financed in part by the Coordenação de Aperfeiçoamento de Pessoal de Nível Superior - Brasil (CAPES) - Finance Code 001 and CNPq and FAPESP. The authors thank the Center for Computational Engineering and Sciences at Unicamp for financial support through the FAPESP/CEPID Grant #2013/08293-7 and Grant #2018/11352-9.